\documentclass[twocolumn,prl,aps,amsfonts]{revtex4}
\usepackage{graphicx}
\usepackage{psfrag, color}
\usepackage{verbatim}

\begin{document}

\title{Large Bychkov-Rashba spin-orbit coupling in high mobility GaN/AlGaN heterostructures}

\author{S. Schmult$^1$, M. J. Manfra$^{1,*}$, A. Punnoose$^{1,2}$, A. M. Sergent$^1$, K. W. Baldwin$^1$, and R. J.
Molnar$^3$}

\address{
  $^1$Bell Laboratories, Lucent Technologies, 700 Mountain
 Avenue, Murray Hill, New Jersey 07974, USA
 \\$^2$Physics Department, University of Wisconsin, 11500 University Ave., Madison, WI 53706, USA
 \\$^3$MIT Lincoln Laboratory, 244 Wood St., Lexington, Massachusetts 02420, USA}

\begin{abstract}
We present low temperature magnetoconductivity measurements of a
density-tunable and high mobility two-dimensional electron gas confined
in the wide bandgap GaN/AlGaN system.  We observed pronounced
anti-localization minima in the low-field conductivity, indicating
the presence of strong spin-orbit coupling.  Density dependent
measurements of magnetoconductivity indicate
that the coupling is mainly due to the Bychkov-Rashba mechanism. In
addition, we have derived a closed-form expression for the
magnetoconductivity, allowing us to extract reliable transport
parameters for our devices. The Rashba spin-orbit coupling constant
is $\alpha_{so}$ $\sim$ 6$\times$ 10$^{-13}$eVm, while the
conduction band spin-orbit splitting energy amounts to $\Delta_{so}$
$\sim$ 0.3meV at n$_e$=1$\times$10$^{16}$m$^{-2}$.

\end{abstract}

\maketitle

GaN has emerged as a leading material for a variety of new device applications, ranging from solid-state, ultra-violet optical sources to high power electronics \cite{Morkoc}.  While the performance of many devices fabricated from GaN has been stunning, several fundamental physical processes remain to be understood.  A prime example is spin-orbit coupling in GaN and its heterostructures.
The burgeoning field of spintronics has invigorated
the study of spin-orbit coupling in semiconducting materials~\cite{datta, zutic}.  To date, much experimental effort has been devoted to narrow bandgap
material systems (InAs, InGaAs, GaAs, etc) as spin-orbit coupling is
expected to be strong in these systems. Conversely, far less
experimental effort has been directed toward wide bandgap systems like GaN in which spin-orbit effects are predicted to be suppressed
by the large fundamental bandgap, E$_g$, and reduced spin-orbit splitting, $\Delta_{0}$, of the valence band at zone center. Indeed,
in the {\bf k$\cdot$p} formalism \cite{cardona}, the bare Rashba
spin-orbit coupling constant for electrons, $\alpha_{0}$, scales as:
$\alpha_{0}$$\sim$ $\Delta_0$/E$^2_g$.  As the value of $\Delta_0$
for GaAs exceeds that of GaN by a factor of 30, it is
reasonable to suspect that spin splitting of the conduction band in
GaN-based heterostructures would be insignificant compared to GaAs
and other narrow gap heterostructures.

Spin-orbit coupling for conduction band electrons in {\it bulk} GaN
was considered by Krishnamurthy \cite{krishnamurthy} who calculated
that the spin relaxation times in bulk GaN should exceed the spin
relaxation time in GaAs by {\it three orders of magnitude}, thus
making GaN an excellent candidate for transport of spin polarized
currents over macroscopic distances.  However, in Ref.
\cite{krishnamurthy} GaN was assumed to have the
zinc-blende lattice structure.  GaN is typically grown in the more
stable wurtzite phase.  It is known that the symmetry of the
underlying crystal has a profound impact on spin-orbit induced
splittings in  the conduction band \cite{Rashba,Lew}.  While the work of
Ref. \cite{krishnamurthy} is suggestive, very few experimental
results for bulk wurtzite GaN have been reported \cite{beschoten} and the impact of spin-orbit coupling on transport in wurtzite GaN/AlGaN heterostructures remains an open question.  A few preliminary experiments have considered spin-orbit coupling for the two-dimensional electron gas (2DEG) \cite{lo,lu,weber,Thillosen} in a narrow parameter space of high density and low mobility.  The influence of spin-orbit coupling in the limit of low 2DEG density and high mobility has not been addressed.  Furthermore, probative experiments in which conductivity is tuned over a broad range, and theoretical analysis specifically tailored to the physics of GaN are needed to understand the mechanisms of spin-orbit coupling and accurately quantify spin-orbit effects in high mobility GaN 2DEGs.

In this Letter we present an analysis of low temperature
magnetoconductivity measurements in a series of high mobility 2DEGs
confined in the wide bandgap GaN/AlGaN system.  Experiments are
conducted with gated Hall bars that allow access to a previously
inaccessable range of low density, $5\times 10^{15}$m$^{-2}\leq$
n$_e\leq1.8\times10^{16}$m$^{-2}$, and very high mobilities 1.4m$^2$/Vs
$\leq\mu\leq8.7$m$^2$/Vs.  We observe non-monotonic behavior in the magnetoconductivity with a pronounced
antilocalization minimum at B$\sim$ 2mT, indicating the presence of
significant spin-orbit coupling.  In addition, we have derived an exact
closed-form expression for the magnetoconductivity, that allows for
the extraction of reliable spin-orbit parameters relevant to our
devices. The relative simplicity of the formula for the magnetoconductivity greatly facilitates data fitting.  Importantly, the magnetic field at
which the magnetoconductivity minimum occurs does not depend
sensitively on electron density.  As we shall show, this result
implies that the Bychkov-Rashba mechanism is the dominant spin-orbit
coupling in our samples.  The extracted Rashba coupling constant
$\alpha_{so}$=6$\times$10$^{-13}$eVm is large, resulting in
spin-split energies ranging from 0.2meV to 0.4meV within the density
range of our experiment.  The value of $\alpha_{so}$ in GaN is
comparable to that seen in the narrower bandgap GaAs
system. 
%Our results demonstrate that the magnitude of spin-orbit%
%splitting, $\Delta_{so}$, in the conduction band of GaN 2DEGs is not%
%solely determined by the size of the spin-orbit gap $\Delta_0$ in%
%the valence band and the bandgap $E_g$.%  
Our findings place severe
constraints on the use of GaN heterostructures for polarized spin
transport, but also suggest that GaN may be implemented in applications where only narrow bandgap materials have been considered previously.

Magnetoconductivity in two dimensions has been studied extensively in the diffusive limit
\cite{Iordanskii,Edelstein}.  The conductivity of a 2DEG in classically weak magnetic fields,
$\sigma(B)$, shows signatures of quantum interference that
depend on the magnetic field and spin-orbit coupling.   Spin
relaxation due to spin-orbit coupling and impurity scattering produces
a positive contribution to the conductivity known as
antilocalization. Magnetic field suppresses this antilocalization.
The functional dependence of $\Delta\sigma(B)$, where $\Delta\sigma(B)=\sigma(B)-\sigma(B=0)$, at small magnetic
fields depends on the relative contributions of dephasing
(characterized by a dephasing rate $1/\tau_{\varphi}$) and the
spin-relaxation rate given by $1/\tau_{so}$.  Strong spin relaxation
compared to dephasing leads to a pronounced minimum in the
magnetoconductivity for excursions away from B=0.  The observation
of an antilocalization minimum in the magnetoconductivity is the
signature of spin-orbit coupling.  

The Bychkov-Rashba interaction in a 2D system can be described by
the  following Hamiltonian \cite{Bychkov}:

\begin{equation}
    H=\frac{\vec{p}}{2m}+\alpha_{so} \vec{\sigma}\cdot( \hat{z} \times \vec{p}).
\end{equation}
Here, $\alpha_{so}$ is the spin-orbit coupling strength, the splitting
energy is $\Delta_{so}$=2$\alpha_{so}$$p_F$ (at the Fermi surface).
In a 2D system with spin-orbit interactions, the dominant spin
relaxation process is typically the D'yakonov-Perel' (DP) mechanism
\cite{Dyakonov}.  It describes the relaxation of the electron spin
in the presence of a spin-splitting, $\Delta_{so}$.  Relaxation
occurs because the direction of the axis of spin precession is tied
to the direction of the electron momentum, which changes randomly
with each collision.  As a result, the net precession after $N$ collisions
is typically $\sqrt{N}\Delta_{so} \tau / \hbar$ in the diffusive
transport regime, where $\tau$ is the mean free time.  Consequently,
the time it takes to randomize the spin is $\tau_{so}\sim
\hbar^2/\Delta_{so}^2\tau$. Expressed in terms of $\alpha_{so}$,
$1/\tau_{so}$=$D(2m\alpha_{so}/\hbar)^2$, where $D=v^2_F\tau/2$ is
the diffusion constant in 2D.

We analyze the measured magnetoconductivity data with an analytical
formula we have derived for $\Delta \sigma(B)$ in the presence of
the Bychkov-Rashba interaction that is valid in the diffusive
regime. The details of our derivation are presented in an upcoming
publication \cite{Alex}. This formulation greatly simplifies
extraction of the transport parameters from experimental data.  The
relevant magnetic field scales are: $B_{so}=\hbar/4eD\tau_{so}$,
$B_{\varphi}=\hbar/4eD\tau_{\varphi}$, and $B_{tr}$=$\hbar/4eD\tau$.
The diffusive limit is defined as $B \ll B_{tr}$.  In this limit,
$\Delta\sigma(B)$ is independent of $B_{tr}$ and the expression for
$\Delta\sigma(B)$ reads:

\begin{widetext}
%\begin{subequations}
\begin{eqnarray}
\Delta\sigma(B)&=&\frac{e^2}{2\pi h}\left[\;\sum_{s=0,\pm 1} u_s
\psi
\left(\frac{1}{2}+b_{\varphi}-v_s\right)-\psi\left(\frac{1}{2}+b_{\varphi}\right)
+\frac{1}{(b_{so}+b_{\varphi})^2-1/4}-2 \ln b_{\varphi}+C\right]~,\\
C&=&-2\ln\left(1+\frac{B_{so}}{B_{\varphi}}\right)-\ln\left(1+\frac{2
B_{so}}{ B_{\varphi}}\right) + \frac{8}{\sqrt{7+ 16
B_{\varphi}/B_{so}}}\; \arccos\left[\frac{2B_{\varphi}/B_{so}-1}
{\sqrt{\left(2B_{\varphi}/B_{so}+3\right)^2-1}}\right].
\end{eqnarray}
\label{eqn:dsigma}
%\end{subequations}
\end{widetext}
The values of $u_s$ and $v_s$ are:
\begin{eqnarray}
v_s&=&{2}\ \delta \cos\left[\theta-\frac{2\pi}{ 3}(1-s)\right]~,\label{eqn:vs}\\
u_s&=&\frac{3 v_s^2+4 b_{so} v_s+(5
b_{so}^2+4b_{so}b_{\varphi}-1)}{\prod_{s'\neq s}(v_s-v_{s'})}~,
\label{eqn:us}
\end{eqnarray}
\label{eqn:usvs}
where the variables $\delta$ and $\theta$ are equal
to:
\begin{eqnarray}
\delta&=&\sqrt{\frac{{1-4 b_{so}b_{\varphi}-b_{so}^2}}{3}}~,\label{eqn:delta}\\
\theta&=&\frac{1}{3}
\arccos\left[-\left(\frac{b_{so}}{\delta}\right)^3\left(1+\frac{2b_{\varphi}}{b_{so}}\right)\right]~.
\label{eqn:theta}
\end{eqnarray}
$\psi(z)$ is the di-gamma function, while
$b_{so}$=$B_{so}/B$  and $b_{\varphi}$=$B_{\varphi}/B$.  The
constant $C$ is such that $\Delta\sigma(0)=0$. Eq.~(2) provides an
analytical, closed-form, solution for the magnetoconductivity in the
presence of the Bychkov-Rashba interaction. The result is
expressed in terms of the two parameters $B_{so}$ and $B_{\varphi}$.
Because, $B_{so}=(m\alpha_{so})^2/e\hbar$, determining $B_{so}$ as a
fitting parameter gives, if the mass is known, directly the value of
the spin-orbit coupling $\alpha_{so}$ defined in Eq.~(1).

The samples used in this study are single interface GaN/AlGaN
heterostructures  grown by plasma-assisted molecular beam epitaxy
(MBE) on (0001) oriented GaN templates.  After an initial 1$\mu$m GaN
buffer layer, a 16nm thick Al$_x$Ga$_{1-x}$N barrier layer (x varies
between 0.08 and 0.12) is grown followed by a 3nm thick GaN capping
layer. The 2DEG is formed at the lower GaN/AlGaN interface without
modulation doping due to the effects of spontaneous and
piezoelectric polarization \cite{ambacher}. Hall bars with 100$\mu$m
width and 2mm length are defined with 14 voltage probes
symmetrically placed along the device. A Ni/Au gate, separated by
50nm of SiO$_2$ from the wafer surface, is used to control the 2DEG
density.  The magnetic field dependent conductivity was measured
with standard low frequency lock-in techniques at T=0.3K.  All
investigated samples show an unambiguous minimum in low field
magnetoconductivity, indicative of strong spin-orbit coupling.

\begin{figure}
\includegraphics[width=0.75\columnwidth]{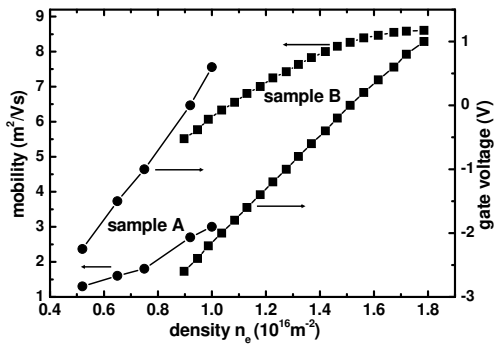}
%\vspace*{.1in}
\caption{Transport parameters of our gated Hall bar devices.  Sample
A  (solid black circles) has lower conductivity, placing it in the
diffusive transport regime.  Sample B (solid black squares) is a
high mobility device.  The combination of the two devices allows
access to a broad range of 2D conductivity.} \label{fig1}
\end{figure}

We discuss in detail two samples with different conductivity (see
Fig.~\ref{fig1}). The first structure, sample A, is designed to have
a mobility and carrier density which place it in the diffusive
limit at low carrier density. With Sample A, we are able to tune the conductivity from the diffusive limit, where our theory is
strictly valid and the extracted parameters are most accurate, to the ballistic regime at higher carrier density.  Our
objective is to extract reliable parameters for sample A in the
diffusive limit and then monitor the evolution of the
magnetoconductivity as the transport moves into the ballistic regime.  Sample B, which has $\mu$ = 8.7m$^2$/Vs at n$_e$=1.8 $\times$10$^{16}$m$^{-2}$, is a very high mobility sample \cite{mike0,mike} which extends our study deep into the
ballistic regime.

\begin{figure}
\includegraphics[width=0.75\columnwidth]{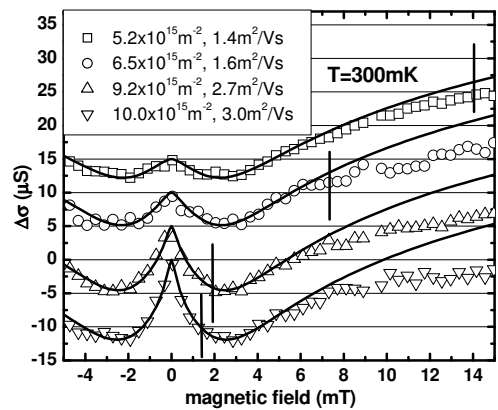}
\caption{Measured magnetoconductivity data (symbols) of sample A for
small magnetic fields over a wide range of density and mobility
(inset). In all cases $\Delta\sigma(B=0) = 0$.  The traces have been
shifted for clearer presentation. The  solid lines denote the fit of
Eq. (2) at each density.  The vertical bars are at the position of
$B_{tr}$, denoting the limit to the diffusive regime.}
\label{fig2}
\vspace{0.3in}
\end{figure}

The experimentally obtained data are plotted as a change in
conductivity $\Delta\sigma(B)  = \sigma(B)-\sigma(B=0)$ and are
shown in Fig.~\ref{fig2} for sample A.  We note that Eq.~(2) predicts a crossover field from negative magnetoconductivity to positive magnetoconductivity at $B \approx B_{so}$.  Several features are evident
in the data. While we change the density by a factor of two and the
conductivity by a factor of four, the field scale at which the
magnetoconductivity minimum occurs does not change within our experimental accuracy. This
fact has two immediate consequences.  As the magnitude of the
spin-orbit coupling constant $\alpha_{so}$ is directly proportional
to $B_{so}$, the data implies that the Bychkov-Rashba coupling strength does
not change as a function of density.  In addition, the lack of field
dependence of the conductivity minimum justifies our neglect of the
linear and cubic Dresselhaus terms in the model Hamiltonian, since
the presence of Dresselhaus coupling dictates that the field scale
$B_{so}$ acquires an explicit density dependence, $B_{so}\sim
\gamma^2n_{e}^2$ \cite{knap,Miller}.

We also see that while the magnetic field position of the
conductivity minimum does not change with conductivity, the
amplitude of the antilocalization effect does.  The antilocalization
induced drop in conductivity is largest at highest conductivity
while it is suppressed at lower conductivity.
Since the field scale of the conductivity minimum, $B_{so}$, is largely constant, the
reduction in the amplitude of the antilocalization behavior can be
attributed to the change in $B_{\varphi}$ as the density and mobility are
reduced.

Using the model presented earlier, the changes to the conductivity
due to quantum interference in the presence of spin-orbit coupling
at each density were calculated (solid lines in Fig.~\ref{fig2}),
allowing for the extraction of $B_{so}$ and $B_{\varphi}$. Also
shown by vertical lines are the values of $B_{tr}$ for each density.
As expected, the deviation of the fits from the data becomes
significant at fields $B\sim B_{tr}$.

\begin{figure}
\includegraphics[width=0.75\columnwidth]{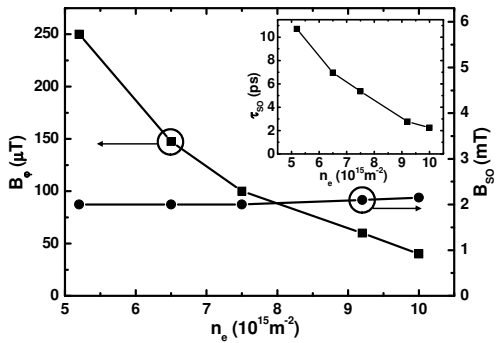}
\caption{The density dependent values of  $B_{so}$ and $B_{\varphi}$
extracted from the fits in Fig.~\ref{fig2}.  While $B_{\varphi}$
decreases by a factor of six with increasing density, $B_{so}$ shows
little density dependence.  The inset shows the evolution of
$\tau_{so}$ as a
function of 2D density.} 
\label{fig3}
\vspace{0.7in}
\end{figure}

The values for $B_{so}$ and $B_{\varphi}$ extracted from the
magnetoconductivity are plotted as a function of 2DEG density in
Fig.~\ref{fig3}.  The field scale $B_{\varphi}$ associated with
dephasing decreases its value as the conductivity increases. The
values for $B_{\varphi}$ correspond to a dephasing time
$\tau_{\phi}$$\sim$100ps in sample A.  As could be expected from
inspection of the raw data, $B_{so}$ remains at $\sim$2mT for the
whole range of densities explored. The corresponding values for the
spin dephasing time $\tau_{so}$ vary from 2ps (high density) to 10ps
(low density).  Using $B_{so}$$\sim$2mT, the spin-orbit coupling
parameter $\alpha_{so}$=6$\times$10$^{-13}$eVm is calculated,
yielding a density dependent splitting, $\Delta_{so}$, of the two
spin subbands at the Fermi edge of 0.2meV - 0.3meV for this sample.

\begin{figure}
\includegraphics[width=0.75\columnwidth]{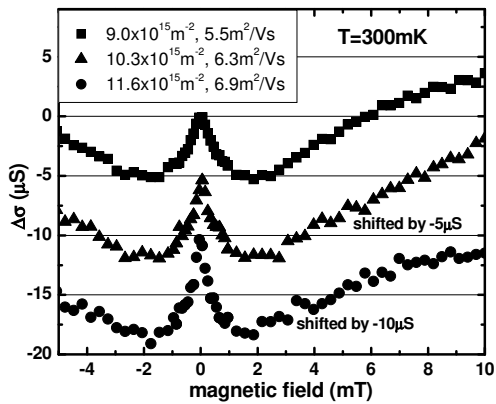}
%\vspace*{.1in}
\caption{Magnetoconductivity  data for sample B.  The high mobility
of sample B dictates that the transport is ballistic.  As with
sample A, the conductivity minimum is situated near B=2mT and does
not depend significantly on 2D density.}
\label{fig4}
\end{figure}

We now turn to sample B.  In sample B, $B_{tr}$ covers the range between
100$\mu$T and 440$\mu$T and is thus significantly smaller than the
field scale $B_{so}$.  It is clear from the forgoing discussion that
our closed form expression for the magnetoconductivity has limited
applicability in this regime.  We therefore do not attempt to fit
the data.  Nevertheless, we still can glean important information
from the raw data.  The magnetoconductivity data for sample B are
presented in Fig.~\ref{fig4}.  The conductivity minimum does not
change position with density and the minimum conductivity is again
near B=2mT, as in sample A.  It follows that the  value of
$\alpha_{so}$ for sample B will be approximately the same as in
sample A.  Indeed, it is not suprising that $B_{so}$ has not changed
significantly in moving from sample A to sample B as the layer
sequence of the two structures is nearly identical.  Given a maximum
density n$_e$=1.8$\times$10$^{16}$m$^{-2}$, a zero field spin
splitting of 0.4meV is calculated for sample B.  It is interesting
to compare our value of spin splitting to the results of Chou {\it
et al}. \cite{HungTao} who examined the zero-field splitting
observed in quantum point contacts (QPC) fabricated on a similar
high mobility GaN heterostructure.  In a QPC built on a GaN 2DEG
with n$_e$=1$\times$10$^{16}$m$^{-2}$ and $\mu$=5.6m$^2$/Vs, Chou
observed a zero-field splitting of 0.39meV, a result which compares
favorably with our value determined from magnetoconductivity
measurements of a GaN 2DEG.

The origin of the large Bychkov-Rashba coupling in GaN heterostructures is
yet to be fully explained.  One possible explanation is found in the
extremely large electric fields at the AlGaN/GaN interface generated
by the polarization discontinuity.  Self-consistent calculations
suggest that the electric field in our GaN heterostructure with
n$_e$=5$\times$10$^{15}$m$^{-2}$, is approximately 10 times the
value in an equivalent density GaAs 2DEG structure \cite{jogai}.
Since $\alpha_{so}$=$\alpha_0$$E$, where $\alpha_0$ is the
fundamental spin-orbit coupling parameter for a particular material
system and $E$ is the value of the electric field at the
heterointerface, it may be that large zero-field spin splitting is
due to the large increase in electric field.  It is also possible that
the fundamental spin-orbit coupling may be enhanced in wurtzite GaN
due to coupling between the conduction band and higher energy
conduction bands.  A calculation of $\alpha_0$ in GaN which accounts
for the wurtzite symmetry and includes remote band effects is still
lacking.  While further study is needed, our experiments clearly
indicate that spin-orbit coupling dramatically influences the
transport properties of high mobility GaN 2DEGs.
\vspace*{10pt}

*Corresponding author: manfra@lucent.com

\end{document}